\documentclass[prl,superscriptaddress,twocolumn,showpacs,preprintnumbers,amsmath,amssymb]{revtex4}

\usepackage{graphicx}
\usepackage{dcolumn}
\usepackage{bm}

\begin{document}

\preprint{Physical Review Letters {\bf 92}, 047002 (2004)}

\title{Gap Structure of the Spin-Triplet Superconductor Sr$_2$RuO$_4$\\ Determined from the Field-Orientation Dependence of Specific Heat}

\author{K. Deguchi}
\email{deguchi@scphys.kyoto-u.ac.jp}
\affiliation{Department of Physics, Graduate School of Science, Kyoto University, Kyoto 606-8502, Japan}

\author{Z. Q. Mao}
\altaffiliation{Present address: Physics Department, Tulane University, 2001 Percival Stern, New Orleans, LA 70118, USA.}
\affiliation{Department of Physics, Graduate School of Science, Kyoto University, Kyoto 606-8502, Japan}

\author{H. Yaguchi}
\affiliation{Department of Physics, Graduate School of Science, Kyoto University, Kyoto 606-8502, Japan}

\author{Y. Maeno}
\affiliation{Department of Physics, Graduate School of Science, Kyoto University, Kyoto 606-8502, Japan}
\affiliation{Kyoto University International Innovation Center, Kyoto 606-8501, Japan}

\date{\today}

\begin{abstract}
We report the field-orientation dependent specific heat of the spin-triplet superconductor Sr$_2$RuO$_4$ under the magnetic field aligned parallel to the RuO$_2$ planes with high accuracy. Below about 0.3 K, striking 4-fold oscillations of the density of states reflecting the superconducting gap structure have been resolved for the first time.  We also obtained strong evidence of multi-band superconductivity and concluded that the superconducting gap in the active band, responsible for the superconducting instability, is modulated with a minimum along the [100] direction.
\end{abstract}

\pacs{74.70.Pq,74.25.Bt,74.25.Op,74.25.Dw}

\maketitle

	Since the discovery of its superconductivity~\cite{discovery}, the layered ruthenate Sr$_2$RuO$_4$ has attracted a keen interest in the physics community~\cite{Review}. The superconductivity of Sr$_2$RuO$_4$ has pronounced unconventional features such as: the invariance of the spin susceptibility across its superconducting (SC) transition temperature $T_{\rm c}$~\cite{NMR,polarised}, appearance of spontaneous internal field~\cite{muSR}, evidence for two-component order parameter~\cite{small} and absence of a Hebel-Slichter peak~\cite{NMR-HS}. These features are coherently understood in terms of spin-triplet superconductivity with the vector order parameter $\bm{d}(\bm{k})=\hat {\bm{z}}{\it \Delta}_{0}(k_x + ik_y)$, representing  the spin state $S_z = 0$ and the orbital wave function with $L_z = +1$, called a chiral {\it p}-wave state.

	The above vector order parameter leads to the gap ${\it \Delta}(\bm{k})={\it \Delta}_{0}(k_x^2 + k_y^2)^{1/2}$, which is isotropic because of the quasi-two dimensionality of the Fermi surface consisting of three cylindrical sheets~\cite{dHvA}. However, a number of experimental results~\cite{Cp,NQR,penetration,kappa,USA} revealed the power-law temperature dependence of quasiparticle (QP) excitations, which suggest lines of nodes or node-like structures in the SC gap. There have been many theoretical attempts (anisotropic {\it p}-wave or {\it f}-wave states) to resolve this controversy~\cite{Gamma,AlphaBeta,Miyake,dxy,dx2-y2,Horizontal}. Although all these models suggest a substantial gap anisotropy, magnetothermal conductivity measurements with the applied field rotated within the RuO$_2$ plane down to 0.35 K revealed little anisotropy~\cite{oscillationT,oscillationI}. To explain those experimental facts as well as the mechanism of the spin-triplet superconductivity, several theories~\cite{HorizontalODS,VerticalODS}, taking the orbital dependent superconductivity (ODS) into account~\cite{Agterberg}, have been proposed. In these models, there are active and passive bands to the superconductivity: the SC instability originates from the active band with a large gap amplitude; pair hopping across active to passive bands leads to a small gap in the passive bands. The gap structure with horizontal lines of nodes~\cite{HorizontalODS} or strong in-plane anisotropy~\cite{VerticalODS} in the passive bands was proposed.

\begin{figure}[b]
\vskip -0.3cm
    \begin{center}
\includegraphics[width=8.6cm,clip]{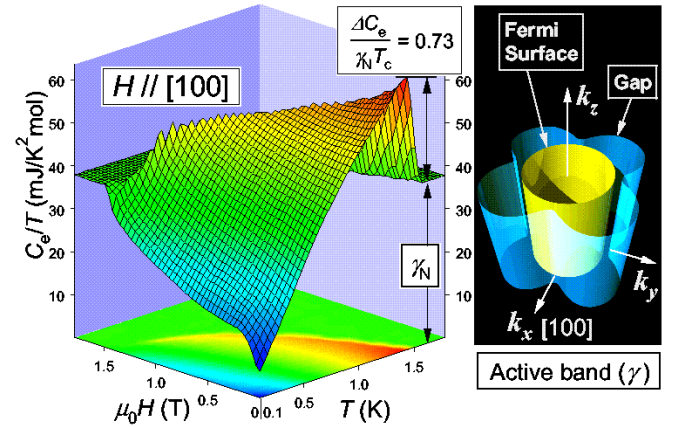}
    \end{center}
\vskip -0.5cm
\caption{\label{fig:3D} Left: Electronic specific heat divided by temperature $C_{\rm e}/T$ for $\bm{H} \parallel [100]$, as a function of field strength and temperature. A contour plot is shown on the bottom $H-T$ plane, with the same color scale as the 3D plot. Right: Superconducting gap structure for the active band  $\gamma$ deduced from the present study, corresponding to $\bm{d}(\bm{k})=\hat {\bm{z}}{\it \Delta}_{0}({\rm sin}ak_x + i{\rm sin}ak_y)$.}
\vskip -0.1cm
\end{figure}

	In order to identify the mechanism of the spin-triplet superconductivity, the determination of the gap structure in the active band is currently of prime importance. The field-orientation dependent specific heat is a direct measure of the QP density of states (DOS) and thus a powerful probe of the SC gap structure~\cite{Vekhter,WonO,Miranovic,Park}. In this Letter, we report high precision experiments of the specific heat as a function of the angle between the crystallographic axes and the magnetic field $\bm{H}$ within the RuO$_2$ plane. We reveal that the SC state of Sr$_2$RuO$_4$ has a band-dependent gap and that the gap of its active SC band has strong in-plane anisotropy with a minimum along the [100] direction, as illustrated in Fig.~\ref{fig:3D}.

	Single crystals of Sr$_2$RuO$_4$ were grown by a floating-zone method in an infrared image furnace~\cite{growth}. After specific-heat measurements on two crystals to confirm the reproducibility of salient characteristics such as a double SC transition~\cite{Double}, the sample with $T_{\rm c} = 1.48$ K, close to the estimated value for an impurity and defect free specimen ($T_{\rm c0}=1.50$ K)~\cite{impurity}, was chosen for detailed study. This crystal was cut and cleaved from the single crystalline rod, to a size of $2.8 \times 4.8 \;{\rm mm}^{2}$ in the $ab$-plane and $0.50 \;{\rm mm}$ along the $c$-axis. The side of the crystal was intentionally misaligned from the [110] axis by $16^{\circ}$. The field-orientation dependence of the specific heat was measured by a relaxation method with a dilution refrigerator.  Since a slight field misalignment causes 2-fold anisotropy of the specific heat due to the large $H_{\rm c2}$ anisotropy $(H_{{\rm c2}\parallel ab}/H_{{\rm c2}\parallel c}\approx 20)$~\cite{Double}, the rotation of the field $\bm{H}$ within the RuO$_2$ plane with high accuracy is very important. For this experiment, we built a measurement system consisting of two orthogonally arranged SC magnets~\cite{VectorMag} to control the polar angle of the field $\bm{H}$. The two SC magnets are installed in a dewar seating on a mechanical rotating stage to control the azimuthal angle. With the dilution refrigerator fixed, we can rotate the field $\bm{H}$ continuously within the RuO$_2$  plane with a misalignment no greater than $0.01^{\circ}$ from the plane.

	The electronic specific heat $C_{\rm e}$ under the in-plane magnetic fields was obtained after subtraction of the phonon contribution with a Debye temperature of 410 K. The left panel of Fig.~\ref{fig:3D} shows $C_{\rm e}/T$ for the [100] field direction, as a function of field and temperature. The figure is constructed from data involving 13 temperature-sweeps and 11 field-sweeps. At low temperatures in zero field, power-law temperature dependence of $C_{\rm e}/T \propto T$ was observed, corresponding to the QPs excited from the line nodes or node-like structure in the gap.

\begin{figure}[t]
    \begin{center}
\includegraphics[width=8cm,clip]{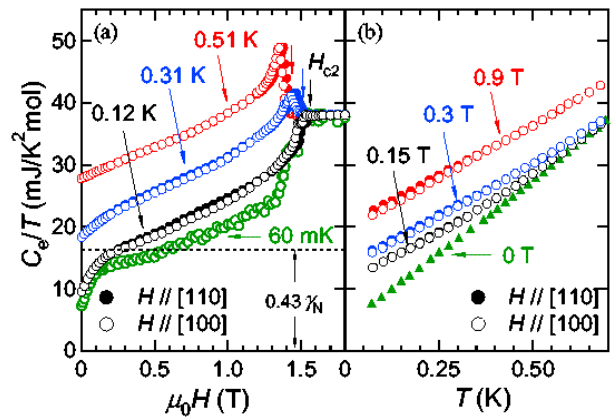}
    \end{center}
\vskip -0.4cm
\caption{\label{fig:CHT} (a), (b) Field and temperature dependence of  $C_{\rm e}/T$ of Sr$_2$RuO$_4$ in magnetic fields parallel to the [100] (open circle) and [110] (closed circle) directions.}
\vskip -0.4cm
\end{figure}

	Now we focus on the field dependence of $C_{\rm e}/T$ at low temperature shown in Fig.~\ref{fig:3D} and Fig.~\ref{fig:CHT} (a). $C_{\rm e}/T$ increases sharply up to about 0.15 T and then gradually for higher fields. This unusual shoulder is naturally explained by the presence of two kinds of gaps~\cite{Cp}. On the basis of the different orbital characters of the three Fermi surfaces ($\alpha, \beta,$ and $\gamma$)~\cite{dHvA}, the gap amplitudes ${\it \Delta}_{\alpha \beta }$ and ${\it \Delta}_{\gamma}$ are expected to be significantly different~\cite{Agterberg}. The normalized DOS of those bands are $\frac{N_{\alpha \beta }}{N_{\rm total}} = 0.43$ and $\frac{N_{\gamma }}{N_{\rm total}} = 0.57$~\cite{Review}. Since the position of the shoulder in $C_{\rm e}/T$ corresponds well with the partial DOS of the $\alpha$ and $\beta$ bands, we conclude that the active band which has a robust SC gap in fields is the $\gamma$-band, mainly derived from the in-plane $d_{xy}$ orbital of Ru $4d$ electrons. Figures~\ref{fig:CHT} (a) and (b) show the field and temperature dependence of $C_{\rm e}/T$ under the in-plane magnetic fields $\bm{H} \parallel [100]$ and $\bm{H} \parallel [110]$ and indicate the existence of a slight in-plane anisotropy.

	In the mixed state, the QP energy spectrum is affected by the Doppler shift $\delta {\it \omega} = \hbar \bm{k} \cdot \bm{v}_s$, where $\bm{v}_s$ is the superfluid velocity around the vortices and $\hbar \bm{k}$ is the QP momentum. This energy shift gives rise to a finite DOS at the Fermi level in the case of $\delta {\it \omega}\geq {\it \Delta}(\bm{k})$~\cite{Volovik}. Since $\bm{v}_s \perp \bm{H}$, $\delta {\it \omega} = 0$ for $\bm{k} \parallel \bm{H}$. Thus the generation of nodal QPs is suppressed for $\bm{H} \parallel$ {\it nodal} directions and yields minima in $C_{\rm e}/T$~\cite{Vekhter,WonO,Miranovic}.
\begin{figure}[b]
\vskip -0.3cm
    \begin{center}
\includegraphics[width=7.2cm,clip]{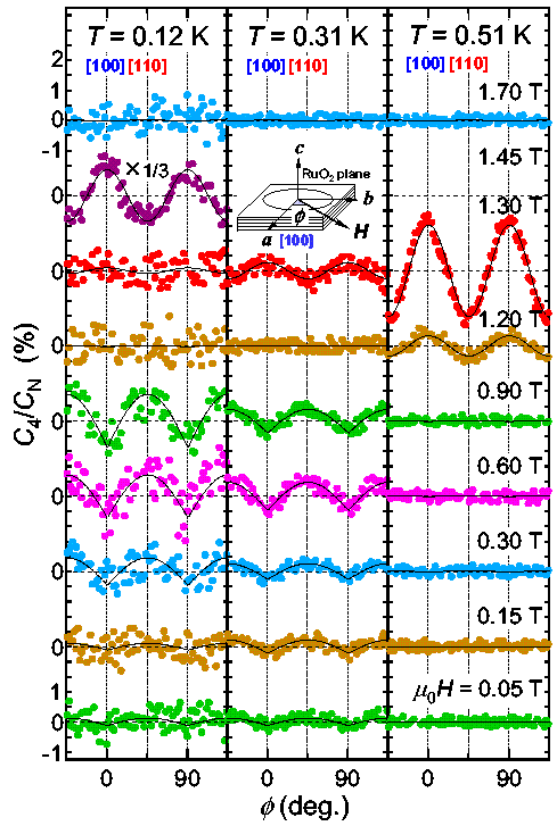}
    \end{center}
\vskip -0.5cm
\caption{\label{fig:OSC}The in-plane field-orientation dependence of normalized 4-fold component of the specific heat at several fields and temperatures. Only 1.45 T data is reduced to 1/3. The solid lines are fits with $f_{4}(\phi)$ given in the text.}
\end{figure}

	Figure~\ref{fig:OSC} shows the field-orientation dependence of the specific heat. The absence of a 2-fold oscillatory component in the raw data guarantees that the in-plane field alignment is accurate during the azimuthal-angle rotation. Thus $C_{\rm e}(T,H,\phi) $ can be decomposed into $\phi$-independent and 4-fold oscillatory terms, where the in-plane azimuthal field angle $\phi$ is defined from the [100] direction: $C_{\rm e}(T,H,\phi) = C_{\rm 0}(T,H) + C_{\rm 4}(T,H,\phi) $. $C_{\rm 4}(T,H,\phi)/C_{\rm N}$ is the normalized angular variation term, where $C_{\rm N}$ is the electronic specific heat in the normal state: $C_{\rm N} = \gamma_{\rm N}T$ with  $\gamma_{\rm N} = 37.8$ mJ/K$^{2}$mol. There is no discernible angular variation in the normal state ($\mu_{0}H = 1.7$ T $> \mu_{0}H_{\rm c2}$); possibilities of angular variation originating from experimental setup or other extrinsic contributions are excluded.

	For fields near $H_{\rm c2}$ ($1.2$ T $\leq \mu_{0}H \leq 1.45$ T), a sinusoidal 4-fold angular variation is observed : $C_{\rm 4}(\phi) \propto f_{\rm 4}(\phi)=-{\rm cos}4\phi$. This is consistent with the in-plane sinusoidal anisotropy of $H_{\rm c2}$ with the maximum in the [110] direction~\cite{Mao,oscillationI}: $C_{{\rm 4}} = \frac{H_{{\rm c2} \parallel [110]}-H_{{\rm c2} \parallel [100]}}{H_{{\rm c2} \parallel [110]}+H_{{\rm c2} \parallel [100]}}\frac{{\rm d}C_{\rm e}}{{\rm d}H}H{\rm cos}4\phi$. Since $H_{\rm c2}$ decreases with increasing $T$, the oscillation amplitude at 1.3 T increases strongly at 0.51 K. For $\mu_{0}H < 1.2$ T, however, a {\it non-sinusoidal} 4-fold angular variation approximated as $C_{\rm 4}(\phi) \propto f_{\rm 4}(\phi)=2|{\rm sin}2\phi| -1$ is observed. Importantly, a phase inversion in $C_{\rm 4}(\phi)$ occurs across about $\mu_{0}H = 1.2$ T: $C_{\rm 4}(\phi)$ takes minima at $\phi = \frac{\pi}{2}n$ ($\phi = \frac{\pi}{4} + \frac{\pi}{2}n$, $n$: integer) for $\mu_{0}H < 1.2$ T  ($\mu_{0}H \geq 1.2$ T), and thus the angular variation for $\mu_{0}H < 1.2$ T cannot be due to the in-plane $H_{\rm c2}$ anisotropy. Therefore we conclude that the {\it non-sinusoidal} 4-fold oscillations originate from the SC gap structure. This result does not contradict the previous measurements of the magnetothermal conductivity down to 0.35 K ~\cite{oscillationT,oscillationI}, which reported little in-plane anisotropy, because these clear oscillations emerge only at lower $T$ $(T/T_{\rm c}\leq 0.2)$.

	For the field range $0.15$ T $< \mu_{0}H < 1.2$ T, where the QPs in the active band $\gamma$ are the dominant source of in-plane anisotropy, we first deduce the existence of a node or gap minimum along the [100] direction, because $C_{\rm e}$ takes a minimum. In addition, we found that the 4-fold oscillations have a {\it non-sinusoidal} form, approximated as $C_{\rm 4}(\phi) \propto 2|{\rm sin}2\phi| -1$, since cusp-like features are clearly seen at the minima ($\phi = \frac{\pi}{2}n$). Strong $k_z$ dependence of the gap function would enhance the QP excitations even if $\bm{H}$ is parallel to the nodal direction, so that the cusp-like features would have been strongly suppressed~\cite{Vekhter}.
\begin{table}[b]
\vskip -0.7cm
\caption{\label{tab:table1}The classified order parameters with the typical gap structures for Sr$_2$RuO$_4$.}
\begin{ruledtabular}
\begin{tabular}{lccl}
 \#&$\bm{d}(\bm{k})$&direction of node or ${\it \Delta}_{\rm min}$&Ref.\\
\hline
1& $\hat {\bm{z}}{\it \Delta}_{0}({\rm sin}ak_x + i{\rm sin}ak_y)$&[100] tiny gap&\cite{Miyake}\\
2& $\hat {\bm{z}}{\it \Delta}_{0}k_xk_y(k_x + ik_y)$&[100] nodes&\cite{dxy}\\
3& $\hat {\bm{z}}{\it \Delta}_{0}(k_x^2-k_y^2)(k_x + ik_y)$&[110] nodes&\cite{dx2-y2}\\
4& $\left\{\begin{array}{cc} \hat {\bm{z}}{\it \Delta}_{0}(k_x + ik_y){\rm cos}ck_z\\\hat {\bm{z}}{\it \Delta}_{0}k_z(k_x + ik_y)^2\end{array}\right.$&horizontal nodes&\cite{Horizontal}\\
\end{tabular}
\end{ruledtabular}
\end{table}

	Most of the proposed gap structures can be classified into four groups as summarized in Table~\ref{tab:table1}.  \#1 and \#2 provide the direction of the gap minima consistent with our observation. To distinguish between \#1 with gap minima and \#2 with nodes, we examine the specific heat jump ${\it \Delta}C_{\rm e}/\gamma_{\rm N}T_{\rm c}$ at $T_{\rm c}$ in zero field. The jump originates mainly from the active band with large ${\it \Delta}$ because of ${\it \Delta}C_{\rm e}/\gamma_{\rm N}T_{\rm c} \propto \partial {\it \Delta}^2/\partial T|_{T_{\rm c}}$. We estimate the contribution of ${\it \Delta}C_{\rm e}/\gamma_{\rm N}T_{\rm c}$ from the active band for the gap structures  \#1 and \#2: ${\it \Delta}C_{\rm e}/\gamma_{\rm N}T_{\rm c} = (1.22$ to $1.07) \times 0.57 = 0.70$ to $0.61$ with the gap minimum (${\it \Delta}_{\rm min}/{\it \Delta}_{\rm max} = 1/2$ to $1/4$)~\cite{Miyake}, while ${\it \Delta}C_{\rm e}/\gamma_{\rm N}T_{\rm c} = 0.75 \times 0.57 = 0.42$ with the line nodes \cite{dxy}. From the experimental result ${\it \Delta}C_{\rm e}/\gamma_{\rm N}T_{\rm c} = 0.73$ in Fig.~\ref{fig:3D} and the estimated additional contribution from the passive bands ${\it \Delta}C_{\rm e}/\gamma_{\rm N}T_{\rm c} \sim 0.04$~\cite{VerticalODS}, $\bm{d}(\bm{k})=\hat {\bm{z}}{\it \Delta}_{0}({\rm sin}ak_x + i{\rm sin}ak_y)$ with the gap minimum is promising for the active band.

\begin{figure}[b]
\vskip -0.2cm
    \begin{center}
\includegraphics[width=8cm,clip]{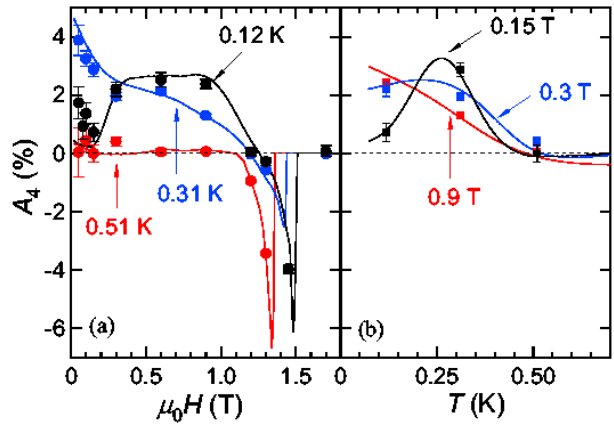}
    \end{center}
\vskip -0.5cm
\caption{\label{fig:C4HT}(a), (b) Field and temperature dependence of the 4-fold anisotropy $A_{\rm 4}$ in the specific heat. The points are evaluated from the fitting to the oscillatory data in Fig.~\ref{fig:OSC}, while the lines from the difference in $C_{\rm e}$ between $\bm{H} \parallel [110]$ and $\bm{H} \parallel [100]$ in Fig.~\ref{fig:CHT}. Two methods yield consistent results.}
\end{figure}

	To facilitate a comparison with theories, although they are presently available only for line-node gaps~\cite{WonO,Miranovic}, we decomposed $C_{\rm e}$ into two parts: $C_{\rm 0}(T,0)$ due to the thermally excited QPs and ${\it \Delta}C_{\rm 0}(T,H)$ due to the field induced QPs, consisting of an isotropic component and a 4-fold anisotropic component $A_{\rm 4}(T,H)$:
$C_{\rm e}(T,H,\phi) = C_{\rm 0}(T,0) + {\it \Delta}C_{\rm 0}(T,H)[1 + A_{\rm 4}(T,H)f_{\rm 4}(\phi)],$
$A_{\rm 4}(T,H)f_{\rm 4}(\phi) = \frac{C_{\rm 4}(T,H,\phi)}{{\it \Delta}C_{\rm 0}(T,H)},$
where $f_{\rm 4}$ was defined previously for low- and high-field ranges. Figures~\ref{fig:C4HT} (a) and (b) show the field and temperature dependence of $A_{\rm 4}$. The field dependence of $A_{\rm 4}$ with a maximum of 4\% anisotropy at 0.31 K shows a monotone decrease from the delocalized-QP dominant region at low fields to the $H_{\rm c2}$-anisotropy dominant region at high fields. The temperature dependence of  $A_{\rm 4}$ with 3\% anisotropy at 0.9 T shows a smooth decrease with increasing temperature. These results are in semi-quantitative agreement with recent theories~\cite{WonO,Miranovic} which predict 4 to 1.5\% anisotropy from gap structures with vertical line nodes.

	In contrast to the theoretical prediction~\cite{Miranovic}, however, at combined low fields ($\mu_{0}H \leq 0.15$ T) and low temperatures ($T \leq 0.3$ K) where QPs on both the active and passive bands are important for the anisotropy, $A_{\rm 4}$ rapidly decreases. This steep reduction of $A_{\rm 4}$ is primarily attributable to the non-zero gap minima ${\it \Delta}_{\rm min}$ of the active $\gamma$ band (\#1). In fact, at the lowest temperatures the 4-fold oscillations are suppressed below a threshold field of about 0.3 T, which should correspond to $\delta {\it \omega} = {\it \Delta}_{\rm min}$. However, even at 0.15 T below the threshold, increasing temperature leads to an increase in $A_{\rm 4}$ only up to about 0.25 K (Fig.~\ref{fig:C4HT} (b)). This behavior is at least in part explained by including contributions of both field and temperature in the QP excitations, but a more quantitative theoretical analysis is needed to examine this possibility.

	Let us finally discuss the roles of the passive bands in the oscillations in $C_{\rm e}$. While the present experimental study has resolved the directions of gap minima in the $\gamma$ band, there still remain two types of possibilities for the passive bands $\alpha$ and $\beta$ to account for the power-law QP excitations at low temperatures: (A) horizontal line nodes (\#4 in Table~\ref{tab:table1})~\cite{HorizontalODS} and (B) vertical gap minima along the [110] directions~\cite{VerticalODS}. The gaps (A) in the passive bands will not contribute to any oscillatory component whether they are fully developed or filled with QPs induced by $H$ and/or $T$. Thus in this case the rather complex $H$ and $T$ dependence of $A_{\rm 4}$ needs to be accounted for solely by the gap structure of the active band. On the other hand, the gaps (B) will give rise to 4-fold oscillations which are out of phase with those originating from the active band, so that the oscillations will be additionally suppressed. Since Fig. 4 (b) shows that $A_{\rm 4}$ at 0.15 T decreases steeply with decreasing temperature in the temperature range where QP excitations strongly reflect the gap structure of the passive bands, the observed steep reduction of $A_{\rm 4}$ may be a consequence of an additional compensation by the passive bands.

	In conclusion, we have for the first time revealed the in-plane anisotropy in the SC gap of the spin-triplet superconductor Sr$_2$RuO$_4$, from the field-orientation dependence of the specific heat at low temperatures. We identified the multi-band superconductivity with the active band $\gamma$, which has a modulated SC gap with a minimum along the [100] direction with little interlayer dispersion. This gap structure is in good correspondence with the $p$-wave order parameter $\bm{d}(\bm{k})=\hat {\bm{z}}{\it \Delta}_{0}({\rm sin}ak_x + i{\rm sin}ak_y)$. For the required nodal structure of the passive bands, the present results may contain decisive information and call for a more quantitative theoretical work to enable the full assignment of the gap structures in all bands.

	We thank N. Kikugawa, K. Machida, T. Nomura, H. Ikeda, Kosaku Yamada, T. Ohmi, M. Sigrist, K. Ishida, A. P. Mackenzie, S. A. Grigera, K. Maki, and T. Ishiguro for stimulating discussions and comments. This work was in part supported by the Grants-in-Aid for Scientific Research from JSPS and MEXT of Japan, by a CREST grant from JST, and by the 21COE program ``Center for Diversity and Universality in Physics'' from MEXT of Japan. K. D. has been supported by JSPS Research Fellowship for Young Scientists.

\end{document}